# Game-Theoretic Approaches for Wireless Communications with Unmanned Aerial Vehicles

Mbazingwa E. Mkiramweni[†], Chungang Yang[†], Jiandong Li[†], Zhu Han[‡]

*Abstract*— **Wireless communications with unmanned aerial vehicles (UAVs) offer a promising solution to provide cost-effective wireless connectivity and extend coverage. In recent years, the area of wireless communications for UAV system design and optimization has been receiving enormous attention from the research community. However, there are still challenges that are far from solved. To cope with those challenges, researchers have been exploring the applicability of game-theoretic approaches. This paper surveys the existing game-theoretic solutions and presents a number of novel solutions, which are designed to optimize energy consumption, enhance network coverage, and improve connectivity in wireless communications with UAVs. We present main game components and the elements they represent in wireless communications with UAVs first and then give a classification of the current used game-theoretic approaches. We identify main problems in wireless communications with UAVs in which game theory has been used to find solutions. We provide a case to show the merits of applying game theory in wireless communication with UAVs. Finally, we discuss shortcomings of the traditional game-theoretic approaches and propose mean field game (MFG) as an appropriate method for solving novel technical problems in massive UAVs networks.**

*Index Terms*— **Game Theory, UAV, Wireless Communications**

## I. INTRODUCTION

REVOLUTIONARY improvements in the technology of unmanned aerial vehicles (UAVs) in the past few decades have led to highly advanced UAVs that come in different shapes, sizes, capabilities, and functions. Nowadays UAVs are mounted with low cost and high performance commercial wireless transceivers such as cellular networks and the IEEE 802.11 [1]. The use of UAVs for wireless communications is one of the most important applications, which is expected to play vital role in the future wireless networks. UAVs equipped with wireless transceivers can be used as relays for transmitting data. They can be deployed as an aerial base station to provide services to the areas without network infrastructure. Also, UAVs can be employed for collecting and delivering data between ground nodes.

There are many advantages of using UAVs for wireless communications over existing network infrastructures. Some of the benefits are as follows.

- **Cost effective:** UAV-aided wireless communication systems are in general less expensive to build compared to a fixed ground base stations. Therefore, UAVs make proper tools for providing cost-effective wireless communications.

- **Swift and easy deployment:** UAVs do not need a runway and move with high velocity. Because of that, they can quickly and efficiently be deployed whenever needed even in a hostile environment. Easy deployment of UAV networks to create an instant communication infrastructure is very useful in emergency situations, such as after a devastating natural disaster.

- **Maneuverability:** Due to the advancement in technology, UAVs are designed with the capability to perform all kind of maneuvers. As a result, UAVs mounted with transceivers can be easily controlled and maneuvered when managing wireless connectivity.

- **Extend coverage:** In a scenario where ground stations cannot communicate due to distance or obstructed line of sight (LoS), UAVs can act as relays to extend the coverage. UAVs can also be used as aerial base stations in areas where there is no, or it is costly to build cellular infrastructure.

- **Enhance connectivity:** Because of their maneuverability, the movement and position of UAVs can be optimized to improve network connectivity. Moreover, connectivity improves significantly when UAV based relays are used compared to the traditional ad-hoc ground networks.

- **Improve performance:** The maneuverability of UAVs through the dynamic adjustment of their states to best suit for the communication environment, and with the aid of short-range LoS of low-altitude UAVs can lead to a significant network performance improvement.

- **Data transmission:** The Internet of Things (IoT) is expected to revolutionize the way we interact with the physical world. There will be massive data transfers

This paper was accepted for publication in the IEEE Wireless Communication Magazine on 04-feb-2018. This work was supported in part by the National Science Foundation of China (61231008); by the Special Financial Grant from the China Postdoctoral Science Foundation (2016T90894); by the Special Financial Grant from the Shaaxi Postdoctoral Science Foundation (154066); by the CETC Key Laboratory of Data Link Technology(CLDL-20162309); by the Natural Science Basic Research Plan in Shaanxi Province of China (2017JZ021); by the ISN02080001; by the 111 Project under Grant B08038; and by the Shaanxi Province Science and Technology Research and Development Program (2011KJXX-40).

[†] M. E. Mkiramweni, C. Yang, and J. Li are with the State Key Lab. of ISN, Xidian University, Xi'an, 710071 China (mbazingwaem@yahoo.co.uk, cgyang@mail.xidian.edu.cn, jdli@mail.xidian.edu.cn).

[‡] Z. Han is with the University of Houston, Houston, TX 77004 USA (email:zhan2@uh.edu), and also with the Department of Computer Science and Engineering, Kyung Hee University, Seoul, South Korea.

between devices that cannot communicate over a long range with a small transmission power. To facilitate and ease data transfer, wireless communications with UAVs can provide a means to collect the IoT data.

In addition, compared to satellites, terrestrial or high altitude platforms, which have longer endurance and wider coverage, UAVs have several advantages. First, they are cost-effective and can swiftly be deployed within a short duration. Second, they are more flexible in reconfiguration and movement, which can enhance network performance. Furthermore, in most scenarios, UAVs can establish a short-distance LoS communication that improves the communication performance.

Deployments of UAVs for wireless communications come with many advantages but also with novel technical challenges. Due to the UAVs high speed, there are rapid changes in link quality, which needs suitable technology for UAV-UAV links and new communication protocol design. Finding optimal height for maximum communication coverage, proper root path for the UAVs and UAV's collision avoidance mechanisms are among the major challenges. As more and more UAVs are expected to be deployed in the future, advanced multi-UAV coordination techniques, resource allocation and interference management schemes, and security mechanisms have to be designed and developed [2].

Currently, researchers have started utilizing game-theoretic approaches for addressing issues in wireless communications with UAVs [3]. Game theory offers many advantages over the conventional methods. First, game theory deals with different problems, where multiple players with contradictory objectives strategically interact with each other in a competition. Therefore, game theory is a natural tool that can be used to characterize the rational behaviors of multiple players. Second, game theory can be used to model interactions between agents, analyze equilibrium and design distributed algorithms. Moreover, game theory has the capability of examining thousands of possible scenarios before taking the best action [4]. In summary, it provides a mathematical framework for analyzing and modeling problems. However, there lacks a survey of game theory for wireless communications with UAVs. This article provides a better understanding of the current research issues and presents game-theoretic solutions, which are designed to optimize wireless communications with UAVs networks. We identify and classify the existing game-theoretic approaches applied in wireless communication with UAVs. We highlight the major technical challenges and then summarize game-theoretic techniques used to solve the challenges. We provide a study case to show the benefits of using game theory. We also discuss interference problem and propose mean field game (MFG) as the best game model that fit the characteristics of massive UAV-aided network characteristics.

The rest of this article is organized as follows. In Section II, we give a brief introduction to game theory and present main game components and the elements they represent in wireless communications with UAVs. We also provide a simple classification of the current used game-theoretic approaches. In section III, we survey the existing game-theoretic solutions which are designed to optimize energy consumption, enhance network coverage, connectivity and network security. We provide a cooperative game case in section IV. We introduce the interference problem, highlight the limitations of traditional game-theoretic models, and introduce MFG as an effective model for solving interference problems in massive UAVs systems in section V, followed by the conclusion in Section VI.

## II. Games in Wireless Communication with UAVs

Game theory is a branch of applied mathematics, which analyzes and describes the strategic interaction among multiple decision-makers. The strategic interaction activities are referred as the games, where each decision-maker chooses the action that gives its own maximum possible outcome at the same time predicting the rational decision taken by the others.

In game theory, entities or individuals who make decisions and perform the actions are referred to as *players*. The moves taken by players in a particular game are the *actions*. The descriptions of how a player can play a game are called *strategies*. They are a complete plan of actions in all possible situations throughout the game. The *payoff*, also known as a reward is what players receive at the end of the game contingent upon the actions of all other players in the game. The payoff is determined by the individual action together with the actions of the competitors.

In the current applications of game theoretic approaches in wireless communications with UAVs, three main components of a game, i.e., players, strategies, and payoffs are represented by different elements of the wireless networks. For instance, UAVs and ground nodes can be represented as players. Beaconing periods scheduling, task servicing, relocating UAVs coordinates, and intruder evasion are examples of the strategies taken by players. Payoffs can be represented by the elements such as a successful encounter with the ground nodes, reduction of energy consumption, performance improvement, coverage maximization, and successful communication of players in the presence of an intruder. All these depend on the specific applications of game theory to the specific technical problems.

In general, game theory models can be classified into two main categories, which are non-cooperative and cooperative games. The existing game-theoretic approaches fall under the two branches as shown in Fig. 2.

### A. Cooperative game

A cooperative game is a structure, in which the agents are allowed to form agreements as a group before choosing their actions. These plans can impact the strategic choices of the players as well as their utilities. Cooperative games are often analyzed through predictions on the coalitions that will be formed, the actions that groups take jointly and the resulting collective payoffs. Examples of cooperative games applied in wireless communication with UAVs are coalition formation game [3] for task allocation and cooperative differential game [9] for network security.

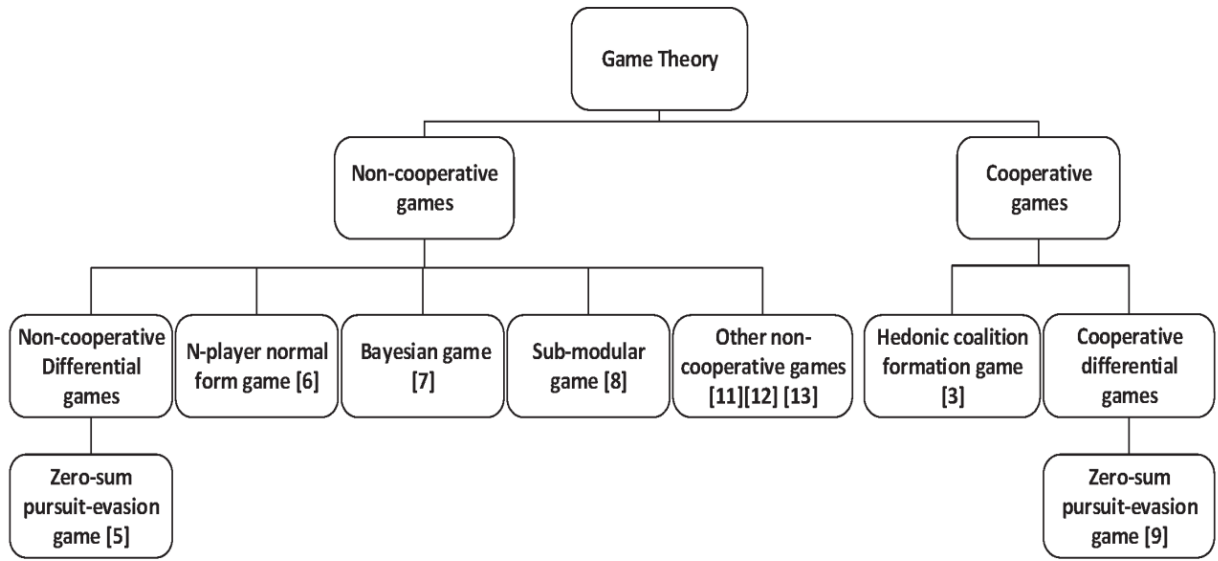

Fig. 1. Classification of current applied game theoretic approaches in wireless communications with UAVs

### B. Non-cooperative games

Contrary to the cooperative game theory, non-cooperative game theory studies strategies among strategically interactive players. The goal of a player is to maximize its payoff (or minimize its cost) by choosing its best strategy individually and rationally. In other words, each player is selfish but rational in a non-cooperative game. The games are mainly applied in power control, distributed resource allocation, coordination of height and position of UAVs, congestion control, UAVs coverage optimization, and spectrum sharing. There exist various kinds of non-cooperative games. Non-cooperative differential games [5], N-player normal-form games [6], Bayesian games [7], and sub-modular games [8] are examples of non-cooperative games applied in solving various technical problems in wireless communications with UAVs networks.

### III. APPLICATIONS OF GAME THEORY IN WIRELESS COMMUNICATIONS WITH UAVS

Game-theoretic properties have been used to solve various types of challenges in wireless communications with UAVs. Non-cooperative games have been used to optimize power and energy, find optimal height and coverage, and perform coordination control of the UAVs. The cooperative game has been applied for tasks allocation and performance efficiency problems. Here, we choose several typical game applications as examples to characterize the rational behaviors, analyze the equilibrium solutions, and design the distributed schemes in wireless communications with UAVs.

### A. Energy Saving and Power Optimization

Energy consumption optimization is one of the major challenges. With the rapid growth of applications with high data rate such as online video streaming and games, more and more energy is consumed. The challenge becomes even more critical when battery-powered UAVs are applied for wireless communications. The operational, performance, and availability duration of UAV are restricted by the limited energy on board. Fig. 1 shows different ways by which a UAV can lose energy. UAVs consume power to overcome its weight so as to stay in the air, as well as for various kinds of motions and maneuvers. Moreover, energy is needed to maintaining inter-UAV communication links such as UAV-UAV, UAV-mobile, UAV-ground control station, and UAV-satellite communication links.

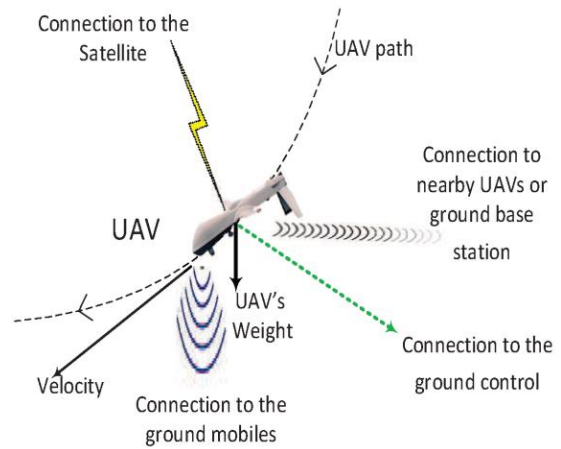

Fig. 2. Different ways by which UAVs consume energy

Improvement in the battery life and energy storage technologies, the use of low-power components and lightweight materials in building drones have not been enough to mitigate the UAV energy problem. From the operational point of view, the problem can be addressed by introducing energy efficiency operations to reduce UAVs unnecessary power consumption during the operation. In addition, researchers have suggested several methods for optimizing UAVs energy consumption such as the UAV circular maneuvering [10]. Such method and others need central authority to control and allocate UAVs to their optimal locations and paths, which increases the energy consumption of UAVs due to the increase of signal overhead. Game theory approaches can allow the UAVs to operate autonomously, therefore minimize power consumption.

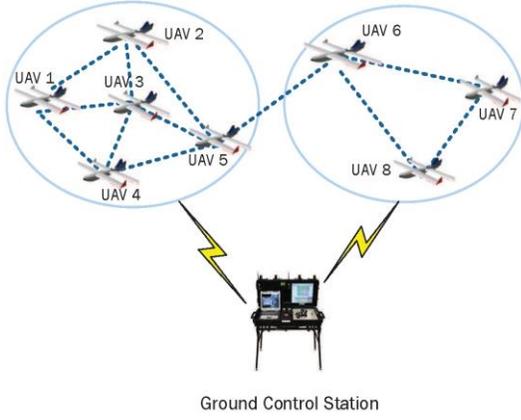

Fig. 4. Ground control station for coordinating UAVs

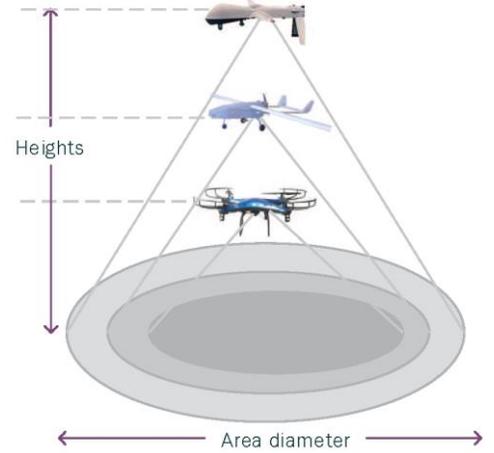

Fig. 3. Area coverage from different heights of UAVs

As an efficient means of power optimization, authors in [8] proposed the use of periodic beaconing for UAVs acting as aerial base stations in a wireless communication system. The authors modeled the beaconing periods scheduling by using a non-cooperative sub-modular game in which the UAVs are considered as players who compete to maximize their coverage probability of the mobiles in the area of interest. Each UAV send a beacon to the mobiles for a specific duration period. A game $G = \{N, \{A_{\{i \in N\}}\}, \{u_{\{i \in N\}}\}\}$ with $N$ being the set of UAVs, $A_i$ set of action and $u_i$ the payoff of UAV$_i$. The payoff function is formulated, which is the difference between the successful encounter rate and energy consumption during the beaconing period. With the provided learning framework, UAVs are allowed to reach the equilibrium, then the existence and uniqueness of the Nash equilibrium are checked, and simulations with different encounter rates are done.

The sub-modular game results show that at the equilibrium point the UAVs efficiently optimize their energy consumption and at the same time maximize the possibility of contacting the mobile user on the ground.

The non-cooperative games in [11] and [12], while designed to find the optimal placement for maximum coverage of the ground mobiles, allows decisions about radio resource and navigation to be made onboard the UAV without reference to other agents. The non-cooperative games use time-stamped broadcasts from the mobiles and UAVs to derive complete information about strategies and payoffs. In this way, the UAVs are allowed to operate with high degree of autonomy and without the need for central planning authority. Therefore, the power required to be assigned to the inter-UAV links, UAV-ground control station link, and the associated communication traffic is minimized.

### B. Optimal Height and Coverage

The altitude at which the UAV is positioned has a significant impact on the power usage, coverage performance, service availability and link reliability over the area that it provides service. Fig. 3 demonstrates how a UAV at different altitudes can have different ground coverage. When the UAVs are properly positioned, the number of UAVs required to provide coverage can be significantly reduced which results in a reduction of resources and time to establish the network. Therefore, determining the optimal height for maximum coverage becomes one of the important research topics.

Game theory concepts have also been applied to find the optimal height and coverage in wireless communication UAVs. The author in [11] uses a non-cooperative game to determine the optimal placement of two UAVs that provide communications to a community of ground mobiles. In this game, the UAVs are the players with two sets of strategies that can be used to cover the ground mobiles uniquely. The first set of strategies contains the choices of next location that the UAVs can take. The other set comprises of the choices to change the altitude during maneuver; this can be ascending, descending or maintaining the altitude depending on the landscape. The aim of UAVs is to choose strategies that will maximize the number of ground mobiles they can support. The UAVs are allowed to find its optimum location based on the location of the mobiles. The results indicate that the UAVs engaged in a competitive game can find its optimum height and significantly better coverage than a single circling UAV. This is because when two UAVs are used, the communications payloads allocated to the support of mobiles is shared between two UAVs, compared to the use of one UAV which will allocate all of its payload power to the support of mobiles.

In [12], three UAVs were used to find their optimum location using a non-cooperative game. In this game, the UAVs are the players. Each player, being mindful of the activity of the other players, strives to maximize their support of mobiles. Their strategies are the next move choices. The payoffs for the players is calculated by a three-stage process. First, by calculating the link budgets of the radio frequency power required to provide the backbone between UAVs. Then the link budgets of all mobiles from each UAV. And the last process is by calculating the other weights such as retaining, adding or dropping mobile. The results indicate that the non-cooperative game offers a useful technique for coordinating the movement of communications UAVs on area coverage missions and increasing the number of UAVs improves the total coverage.

The use of game theory in finding the optimal height and coverage permits the UAVs to make decisions about radio resources and navigation on board without the need for a central planning agency. The UAV-ground control communication overhead is minimized, saving power which is directed to providing coverage. The major drawback of these games is that they do not account for the interference between UAVs and do not take into consideration UAV-UAV connectivity quality.

### C. Coordination of UAVs

In a scenario where many UAVs are deployed, it is expected that more complex systems of selfish individual UAVs with different missions and coalitions will be formed. In order to accomplish their missions and maximize the number of mobiles that UAVs can support with the maximum throughput and minimum energy, each group or individual UAV will seek to find its optimal position and height, move with different speed at different heights and possibly in different directions. In such a scenario, better coordination mechanisms for sharing data have to be designed. Currently, UAVs coordination is primarily done by the ground control station as depicted in Fig. 4, where each UAV or a group of UAVs require a ground controller. Better coordination will ensure the maximum coverage, better quality of service and good network connectivity.

Non-cooperative game theory and evolutionary algorithms are among the technical approaches for coordinating UAVs. Authors in [13] compare the two methods for optimally locating the UAVs. A non-cooperative game mechanism considers UAVs as players that select the best actions to maximize their payoffs. Payoffs are the number of mobile users that the UAVs provide coverage. In this game the players have perfect information, therefore can determine other players strategies and payoffs. The players will choose the best strategy by changing their positions or heights until they attain equilibrium. At the Nash equilibrium, each UAV is expected to know the strategy selected by the others and determine which maneuver will undertake next. In the evolutionary algorithm approach, a master UAV which is responsible for running the algorithm and distributing the generated solutions is selected. The generated solutions consist of a set of flying instructions that allow maneuvering for each UAVs. Using predicted positions which based on recent coordination data of both mobiles and UAVs, the evaluation of the survival of the fittest UAV is estimated. During the evaluation, the resulting individual UAV coverage is estimated.

The results in [13] show that the evolutionary algorithms converge on the optimal solution quicker than the non-cooperative game due to their flexibility. Nevertheless, the non-cooperative game has less risk of flying UAVs out of the operation area because of the use of more conservative flying maneuvers. Moreover, the UAVs driven by the non-cooperative game are found to distribute the load equally among themselves because the coverage is evenly divided among the UAVs. The distribution of load has led to a mission with power balanced which can offer benefits when managing interference of radio frequency between the platforms. Another advantage of using the game theory for coordination is that UAVs becomes autonomously and adaptively.

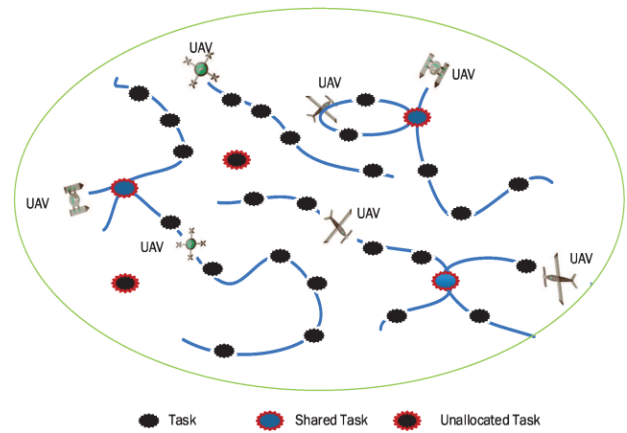

Fig. 5. Different UAVs performing various tasks in a wireless network

### D. Task Allocation and Performance Efficiency Optimization

UAVs in wireless communications can be assigned to perform different tasks such as collecting, transferring and delivering of data, and also monitoring of randomly located sites such as oil, gas, and water pipelines. Fig. 5 shows a scenario where different UAVs are assigned to perform various tasks in a network, and some tasks are performed by more than one UAV while others are not allocated to any of the UAV. With the continuous increase in traffic, applications, and services in the large-scale wireless systems, as a means to enhance performance, mechanisms for organizing and allocating tasks among UAVs are needed. Due to system complexity and dynamic nature of nodes and UAVs, centralized pre-generating and assigning tasks as a means of allocation tasks for the UAVs will not be practical. Therefore, the designed mechanisms have to allow self-organizing and self-adapting of the network while performing and share randomly generated tasks allocation among UAVs. Researchers have started to use game theory to address this challenge.

The authors in [3] modeled a hedonic coalition formation game between the agents and the tasks that interact to form disjoint coalitions. In this game, agents and tasks are the players. Agents act as collectors that collect the packets continuously from randomly located nodes or relays that wirelessly transmit the collected packets to a centralized receiver. Tasks represent a data source such as a group of mobile devices that require servicing, data generated from video surveillance or any other source of packet data that is to be received and transmitted by the UAV. The cooperative groups containing a number of tasks and UAVs are formed according to the proposed algorithm. The UAVs belonging to the same coalition can arrange themselves onto relays and collectors. To optimize throughput and delay, the players can join or leave the coalition based on their preference. The payoff takes into account the benefits received from servicing a task, in terms of effective throughput as well as the cost in terms of delay incurred from the time needed for servicing all the tasks in a coalition. The payoff is then equally divided among the players. The results of the proposed hedonic coalition formation algorithm have shown that the UAVs and the tasks can self-organize in forming coalitions and therefore improve average performance of UAVs compared to the scheme that allocates task equally among the UAVs.

*E.   UAVs Network Security*

UAV-aided networks are exposed to attacks that could cause an enormous amount of loss regarding money, reputation and data confidentiality. Attackers can jam the communications between the UAV and controller, take control of a targeted UAV and launch another kind of attacks such as GPS spoofing. Therefore, an efficient security mechanism to protect such network against attackers is essential. Because of its ability to deal problems where multiple entities with contradictory objectives are involved, game theory technique can be used to model and analyze wireless network security issues.

To protect the UAV-aided network against external and internal intruders, authors in [7] designed the intrusion detection system (IDS) and the intrusion ejection system (IES) for monitoring the network and ejection of the node that is anticipated to instigate an attack. Authors formulated two security game problems as a Bayesian game. The first game involves IDS and attackers as players who possess set of strategies, and a defined set of profit gained according to the strategies. Then the optimal solution defined as Bayesian Nash equilibrium (BNE) is determined, in which the attacker launches a malicious act and IDS activates its monitoring process before and during the malicious act. The second game is between the IES and suspicious nodes. In this game, when the node is categorized as an attacker by IES, it will first change its future state to either transitory or permanent. A BNE optimal solution is determined to check if the suspected node continues to be an attacker in future state, and if so, the IES will eject this suspected node before the attack. The performance of this approach based on the Nash Equilibrium concept has shown to incur a low overhead to detect lethal attacks with a high accuracy.

To compute optimal strategies for a team of UAVs evading the attack of an aerial jammer on the communication channel, the authors in [9] proposed a differential game theoretic approach. The model uses three nodes, receiver, transmitter and the attacker that is attempting to jam the communication channel between the transmitter and receiver by sending a high power noise at the same frequency. Two problems are formulated; the first one by considering that UAVs are initially not communicating in the presence of a jammer. In this first problem, the UAVs seeks to minimize the time for which communication remains jammed while the jammer aims to maximize the time for which it can jam the communication between UAVs. The second problem considers a situation in which UAVs are communicating in the presence of a jammer, which aims to minimize the time it can jam the communication channel between UAVs. Meanwhile, each UAV is trying to maximize the time for which they maintain the operation of the communication link between them. The problem was formulated as a zero-sum pursuit-evasion game and the necessary conditions to arrive at the equations governing the saddle point strategies of the players were derived. The cost function was formulated as the termination time of the game. The game renders a possible way of protecting the communication channel to malicious attacks from aerial intruders flying in the vicinity.

## IV.   A COOPERATIVE BEACONING SCHEDULING STRATEGY BASED ON NASH BARGAINING SOLUTION: A CASE STUDY

The current applied game theoretic approaches have shown promising potentials for solving problems in wireless communications with UAVs [7]. To verify the merits of carrying UAV-aided communications based on game theory, we present a cooperative beaconing periods scheduling game between two UAVs. Motivated by the scenario presented in [8], where two UAVs acting as aerial base stations provide network coverage to mobile users in a geographical area. In this scenario, to encounter a mobile user, the UAVs send beacons announcing their presence. Two UAVs seek to minimize their energy consumption by adopting a periodical probing strategy. Different from [8] where a sub-modular, non-cooperative game was used, we apply cooperative strategy based on the Nash bargaining solution (NBS) to find the optimal probing period to maximize UAVs encounter probability. More details on the NBS are given in [14]. The NBS strategy can achieve an optimal solution while maintaining fairness among UAVs.

In this case study, the two-player bargaining problem is modeled. We first defined players' utility $U_i$, which the UAV achieves when the beaconing period duration is allocated to it, and the minimum utility $\overline{U_i}$, which a UAV obtains when no agreement is reached in the bargaining process. Utility $U_i$ is defined as the difference between the probability of the UAVs' first encounter with the mobile and the cost involved. The cost is the energy consumed associated with sending a beacon and switching between states (i.e., probing/idle states). In this game, if a player receives a utility value less than $\overline{U_i}$, will quit the cooperation. Then a two-player NBS function is formulated as the product of the difference between the defined utilities, i.e., $(U_1 - \overline{U_1}) \times (U_2 - \overline{U_2})$. The optimal beaconing periods of UAVs that give maximum values of the NBS function at various encounter rates are calculated. Finally, we provide simulation results to illustrate the effectiveness of the proposed NBS.

The optimal beaconing periods obtained using NBS for various values of encounter rates at three different cycle

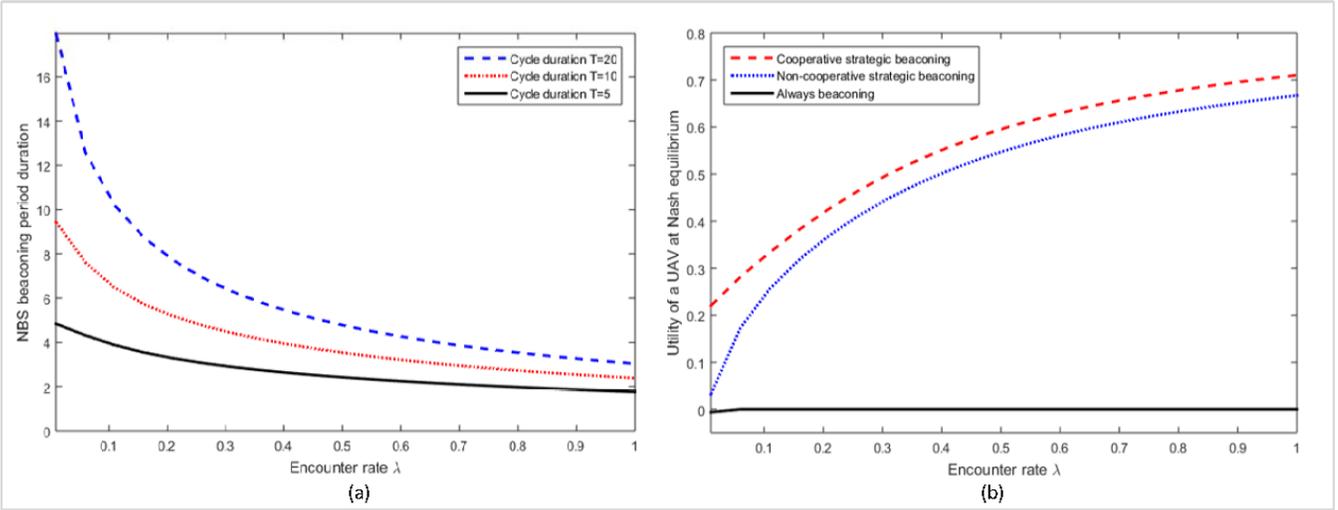

Fig. 6. (a) Beaconing period durations at equilibrium for different values of encounter rates using Nash bargain solution. (b) The utility achieved by the UAV using the cooperative, non-cooperative, and always-beaconing strategies

durations can be seen in Fig. 6 (a). For different probing/idle cycling periods T, for example when T=20, T=10, and T=5, the optimal beaconing period durations start at high values for low values of encounter rates and decreases as the encounter rates increases. The results demonstrate that, when a UAV has a low probability of encountering a mobile, it will require longer probing period durations and vice versa. In addition, at Nash equilibrium, longer beaconing durations will be assigned to the UAVs with higher probing/idle durations compared to the ones with low probing/idle durations. Fig. 6 (b) presents the comparison of utilities that a UAV can achieve when applying three different strategies. The cooperative strategy can achieve higher payoff compared to the always-beaconing and non-cooperative strategies. For the UAV that uses always-probing strategy, there is a linear trade-off between probing duration and energy consumption. As the probing duration increases, the probability of encountering a mobile and associated energy consumption increases as well. However, the UAVs that employ game theory can minimize the energy consumption involved through maximizing their payoff by obtaining the optimal probing period durations. Moreover, it can be seen in Fig. 6 (b), at the optimal solution, the NBS strategy receives higher reward compared to the non-cooperative strategy, which indicates that the cooperative strategy is more efficient in serving energy. This use case shows the potentials of game-theoretic approaches in wireless communication with UAVs.

## V. NOVEL GAME-THEORETIC APPROACHES AND FUTURE RESEARCH DIRECTIONS

In addition to the mentioned novel technical challenges, there exist other problems. In particular, here we concentrate on the interference problems in the massive UAVs communications, which calls for novel game-theoretic approaches.

### A. Interference Management for Massive UAVs

Severe interference is one of the major challenges in designing and deployment of massive UAVs for communication systems. Interference has always been a key threat to wireless communications. Because self-organization networks of UAVs use the sharing wireless channel for transmitting information, they are facing complex wireless interference. In addition, the increasing demands of spectral efficiency, data rate, degrees of freedom, and network capacity due to the exponential growth of mobile users, it is expected that massive UAVs for wireless communications will be deployed. The large number of UAVs will intensify the interference problem, which can severely affect the performance of a wireless network. Moreover, since wireless communications in the air have line-of-sight propagation characteristics, the interference problem becomes more severe. Due to the mobility nature of UAVs, the use of wireless backhauls as well as the lack of centralized control, interference coordination among UAVs is more challenging than in terrestrial cellular systems. Therefore, special techniques for interference management specifically designed for UAV-assisted wireless communication networks are needed.

In a typical wireless communication with UAVs networks, there exist different types of interference as follows.

- Interference from neighboring ground control stations (GCSs). This kind of interference is caused by signals from neighboring GCSs or other devices that use the same radio frequency (e.g., 2.4 GHz) interfering UAV-GCS communication channel.

- Interference among neighboring communicating UAVs. Communicating UAVs will send or receive signals to and from the established communication links; meanwhile, they will receive and send signals to unintended neighboring UAVs, which cause mutual interference.

- UAVs inter-cell interference. When UAVs mounted with transceivers are applied as base stations, they will broadcast radio signals with specific bands that will be detected by mobile users. If adjacent UAVs broadcast using the same bands, they will cause interference at the overlapping edges of the coverage areas. In summary, a user at the overlapping edges of adjacent cells, while communicating through one UAV cell, will experience signal interference from the adjacent UAV cell. Inter-cell interference coordination in UAV-based networks remains a challenge even when the channel frequency reuse technique is applied due to the mobility nature of UAVs.

- UAVs self-interference. Another major source of interference on UAVs are the components installed on UAV. For instance, for small size UAVs, the global navigation satellite system antennas are installed in close proximity to other electronic systems.

- Interference from an intruder. Since UAVs operate in the sky, they are more vulnerable to interference from the jammer due to the fact that in the sky signals can propagate over longer distances than they would on the ground. Different from the ground where signals can be hindered by mountains, trees, buildings and other obstacles, in the sky the intruders' signal have a potential to reach further unhindered.

Different from the ground-based communication infrastructures, in the air, interfered UAVs can result into serious consequences. Moreover, interference seriously affects the network transmission capacity and performance. Reducing interference can improve UAV network security, minimize conflict and retransmission of signals, thus reducing the waste of energy and improving the network throughput. The traditional game models applied in UAV-aided wireless communications networks are not sufficient when analyzing and solving interference problem in a massive UAVs communication networks due to, but not limited to the following reasons.

- The immense signaling and communication overhead caused by information exchange among different UAVs (players) in the network.

- In large-scale multi-agent networks, the traditional approaches are impractical. In these games, every player should collect the information of all other players, which can result in complex algorithms that are considerably harder to solve when dealing with massive networks.

- As the number of UAVs increases, the size of the payoff matrix increases, therefore the time required for the game to reach equilibrium also increases.

Therefore, it becomes necessary to rethink of the different analytical models of tackling the problem in future massive, dynamic and complex UAVs wireless networks and move toward models that fit the characteristics of the network more appropriately. Thus, we propose to use the concepts of mean field game (MFG) [15] to study the characteristics of interference and design interference aware management scheme among UAVs in large wireless communication with UAVs networks.

### B. Mean Field Game for Interference Management in Massive UAV Networks

There is a growing interest for various enterprises and government institution in using UAVs for different purposes. The existing and expected uses of UAVs and huge investment in the drone industry indicates that the future where hundreds if not thousands of drone will be operating in the air is near. Massive UAVs will also be used as part of the coming of 5G networks, to enhance capacity, increase data rates, minimize latency, improve connectivity for a massive number of devices, reduced energy and cost, and boost the quality of experience (QoE).

Therefore, MFG will be an appropriate method for addressing different challenges in such massive UAV networks. This is because in MFG the mean field value of space-time dynamics of context approaches the real value as the number of players increase. In addition, in the MFG framework, the concept of mean field can characterize the space-time dynamics of context, for example, two-dimensional interference and energy states, which helps the generic player to make an optimal decision responding only to the mean field, instead of the strategies of all the other players. Another unique advantage of the introduced novel game is that MFG-based decision-making is distributed with less signaling overhead.

By definition, MFGs are a special form of differential games, where each player has a state, a set of actions, and a control policy. The control policy maps every state into an action over a pre-defined period of time. Instead of modeling each players' interaction with every other player, MFG models an individuals' interaction with the effect of the collective behavior of all the players. Thus, at each step of the game, the mean field is simply the fraction of players at every state. Contrarily to traditional games, MFG can be used to model and analyze the interaction between individual and collective behaviors of a large number of rational entities [15].

In finding a solution to the interference problem, it is necessary to learn about the behavior of the interference. We intend to learn the characteristics of interference that communicating UAVs experience from the neighboring GCSs, mobile devices, and neighboring communicating UAV. We will use the MFG to study and develop a distributed interference-aware scheme. In our model, we will consider a massive UAV-aided wireless network in which UAVs communicate with GCS, other UAVs and provide services to the mobiles. In this model, we will focus on the interference that a UAV communicating with GCS experiences from neighboring GCSs. The mean field approximation method will be used to approach the aggregate interference from other links to UAV-GCS communication channel and get the corresponding cost function.

UAV-GCS links representing a communication pairs are considered as players. They are rational policymakers whose number is arbitrarily large and even goes to infinite. The actions are the possible transmit powers corresponding to UAV-GCS pairs. Each transmitter determines the power it transmits at any time to minimize the cost function. State space of player is defined as the interference introduced by other links to UAV-GCS link and the interference from UAV-GCS link to other links. A control policy is used to minimize the average cost over the time interval in one-dimensional states. The cost function for interference management in massive UAV communications will be formulated by considering both the achieved signal-to-interference-plus-noise ratio (SINR) performance and the transmit power of the UAV-GCS pair.

The interaction of players with the mean field is modeled by a Hamilton-Jacobi-Bellman (HJB) equation, and the movement of the mean field function according to the players' actions is defined by a Fokker-Planck-Kolmogorov (FPK) equation. The corresponding HJB and FPK equations are derived and solved to obtain the mean field equilibrium (MFE). The finite difference method and Lagrange relaxation method will be used to solve the FPK and HJB equations respectively.

### C. Other Research and Implementation

Due to the promising development of IoT and mobile Internet, there exist novel communication requirements of ultra-reliable low latency communications (uRLLC), massive machine-type communication (eMTC), and enhanced mobile broadband (eMBB). Meanwhile, UAVs are also expected to be used to assist vehicular ad-hoc networks (VANETs) and wireless sensor networks (WNS). It is necessary to study the characteristics and advantages of deploying UAVs to assist these networks. Novel game-theoretic approaches can find wider applications.

In the future, it is also necessary to design and implement a prototype of the UAVs-aided wireless communications network, and then evaluate the feasibility of game-theoretic approaches. Then, a series of experiments will be practically conducted and the results can be collected and analyzed to further understand the pros and cons of game-theoretic approaches.

## VI. CONCLUSION

Game theory offers a suitable tool that can be used to effectively model the interaction between autonomous wireless communication UAVs. Game theory always leads to sophisticated distributed decision processes. This article summarized the typical advantages and typical challenges of UAVs communications, where we concentrated on the recent applications of game theory for wireless communications with UAVs. We presented a classification of the existing game approaches to various novel communication challenges with UAVs. For each game-theoretic application, we characterized the rational behavior, analyzed equilibrium solutions, and presented the distributed schemes of the UAVs. Finally, we simulated a use case to show the potential benefits of game theory. We highlighted the interference management challenge, presented the novel mean field game-theoretic approaches, and lastly, looked forward the future of game theory for wireless communications with UAVs.